\documentclass{aa}
\usepackage{psfig}
\begin{document}
\thesaurus{(02.01.2; 02.13.1; 08.14.1; 08.13.1; 13.25.5)}
\title {ROSAT X-ray sources and 
exponential field decay in isolated neutron stars}

\author{S. B. Popov  \and
  M. E. Prokhorov
}
\offprints{S. B. Popov: polar@xray.sai.msu.su}

\institute{
Sternberg Astronomical Institute,
 Universitetskii pr.13, 119899, Moscow, Russia\\
e-mail: polar@xray.sai.msu.su; mystery@sai.msu.su
}

\date{}

\authorrunning{S. Popov \& M. Prokhorov}
\titlerunning{Field decay in isolated neutron stars}

\maketitle

\begin{abstract}
 In this paper we semianalyticaly evaluate influence of the
exponential decay of magnetic field on the fate of 
isolated neutron stars. 
The fact of ROSAT observations of several X-ray sources, which can be
accreting old isolated neutron stars gives us an opportunity to
put some limits on the parameters of the exponential decay.

We argue, that, if most part of neutron stars 
have approximately the same decay and initial parameters, 
then the combinations of the bottom magnetic momentum, $\mu_b$, 
in the range $\sim 10^{28}-10^{29.5} \, {\rm G}\, {\rm cm}^3$ 
and characteristic time scale, $t_d$, in the range
$\sim 10^7-10^8\, {\rm yrs}$ for standard 
initial magnetic momentum, $\mu_0=10^{30} \, {\rm G}\, {\rm cm}^3$,
can be excluded, because for that sets of parameters
neutron stars never come to the stage when accretion of the interstellar
medium on their surfaces is possible even for low velocity of neutron stars
and relatively high density of the interstellar medium. 
The region of excluded parameters increases with $\mu_0$ decreasing.
\end{abstract}


\section{Introduction}

Evolution of neutron stars (NSs) can be called ``magneto--rotational'',
because all 
main astrophysical manifestations of these objects are determined by
their periods and magnetic fields.

Four main regimes, as described for example in Lipunov (1992), are possible
for isolated NSs: ejector, when a star represents a radio pulsar, or a dead
pulsar, spinning down due to magneto--dipole radiation; propeller, when
surrounding captured matter cannot penetrate through the centrifugal
barrier; accretor, when matter can reach the surface, and the NSs appear as an
X-ray source; and georotator, when gravitation becomes insignificant,
because magnetic pressure  dominates everywhere over the gravitational pull,
and geo-like magnetosphere is formed.

Magnetic field decay in NSs is an uncertain subject. Many models were
suggested (see, for example, Ding et al., 1993; Jahan Miri  \& 
Bhattacharya, 1994 ). The only strong observational 
result is that for the exponential (or nearly exponential) decay
characteristic time scale, $t_d$, is longer than $\sim 10^7 \, {\rm yrs}$, 
because no
expected effects of decay are observed in radio pulsars (Lyne et al. 1998).

Field decay was used in the case of old accreting isolated NSs
by Konenkov \& Popov (1997) and Wang (1997) to explain
properties of the source RX J0720-3125. And it seems, that if this source
really is an accreting old NS and if it was born as  a normal radio pulsar
(with small period, $\ll 1$ s,  and high magnetic field, $\sim 10^{12}$ G), 
its properties can be explained only if decay
is working (or if this source was born with such unusual parameters).
Recently 
the influence of the field decay in isolated NSs 
was investigated in Colpi et al. (1998) and Livio et al. (1998).
An attempt to include field decay into population synthesis of isolated
NSs was made in Popov et al. (1999). 

Here we try to put some
limits on the parameters of the exponential field decay, assuming, that 
old isolated NSs are observed as accreting X-ray sources  
(Walter, Wolk \& Ne\"uhauser 1996;
Haberl et al. 1996, 1998; Ne\"uhauser \& Tr\"umper 1999; Schwope et al. 1999),
neglecting the possibility, that all of them can be highly magnetized NSs,
``magnetars'' (see a brief discussion on this assumption in Popov et al., 1999).

\section{Calculations and results}

 The main idea of our work 
is to calculate the ejector time, $t_E$, i.e. a time interval spent by a NS
on the ejector stage, for different parameters of the field decay and
standard assumptions on the initial parameters of a NS,
and to compare that time with the Hubble time, $t_H$. 

For constant field $t_E$ monotonically depends upon NS's velocity and ISM
density:

\begin{equation}
t_E(\mu=const)\sim 10^9 \mu_{30}^{-1}n^{-1/2}\, v_{10} \,{\rm yrs}.
\end{equation}
If $t_E$ for decaying field
for some sets of parameters is greater than
$t_H \sim 10^{10} \, {\rm yrs}$ even for relatively high concentration 
of interstellar medium (ISM),  $\sim 1 \, {\rm cm}^{-3}$, 
and  small spatial velocity of a NS, $\sim 10 \, {\rm km}\, {\rm s}^{-1}$
(this velocity is about the sound velocity in the ISM), 
than for that parameters of the decay, $t_d$ and $\mu_b$, 
no accreting NSs would be observed, and such sets can be excluded as
progenitors of old accreting isolated NSs. In addition, if we
assume, that most part of NSs evolve with similar $t_d$ and $\mu_b$ and are
born with some typical initial period, $p_0$, and initial magnetic momentum,
$\mu_0$ (that is supported by radio pulsars observations), then such sets
of parameters should be totally 
excluded from models of the magnetic field decay.

\begin{figure}
\vbox{\psfig{figure=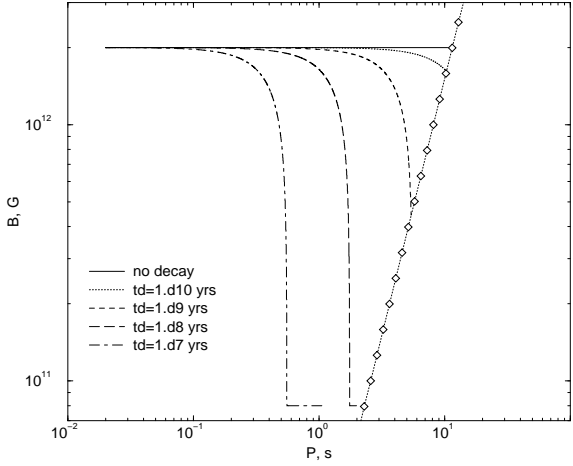,width=9.0cm,angle=-90}}
\caption[]{Tracks on P-B diagram. Tracks are plotted
for bottom polar magnetic field
$8\cdot 10^{10}\, {\rm G}$, initial polar field $2\cdot 10^{12}\, {\rm G}$,
NS velocity $10 \,
{\rm km} {\rm s}^{-1}$, ISM density $1\, {\rm cm}^{-3}$ and different $t_d$.
The final points for different tracks are corresponding to 
the following moments: for $t_d=10^7$ yrs to $\sim 10^{10}$ yrs;
for $t_d=10^8$ yrs to $\sim 10^{10}$ yrs; for $t_d=10^9$ yrs to $\sim 
1.5\cdot 10^9$ yrs; for $t_d=10^{10}$ yrs to $\sim 2 \cdot 10^9$ yrs.
Here we used $p_0=0.020$ s. The line with diamonds
shows the ejector period, $p_E$.}
\end{figure}

\begin{figure}
\vbox{\psfig{figure=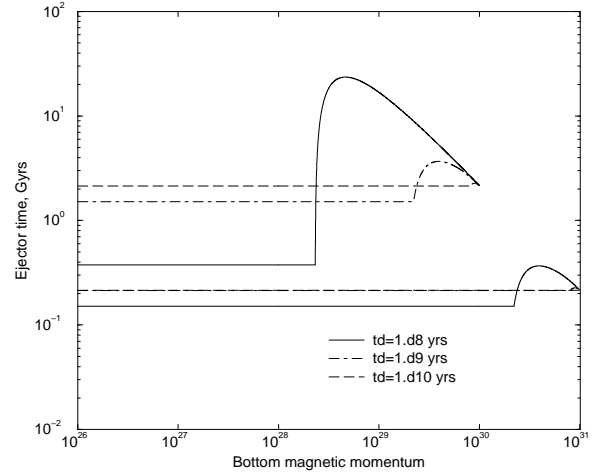,width=9.0cm,angle=-90}}
\caption{ Ejector time, $t_E$, (in billion years) vs. bottom magnetic
momentum.
Curves are shown for two values of the initial momentum:
$10^{30} {\rm G} \, {\rm cm}^3$ (upper curves)
and $10^{31} {\rm G} \, {\rm cm}^3$. }
\end{figure} 

As the first approximation 
only $t_E$ can be taken into account, because the time interval,
spent on the propeller
stage, $t_P$, is uncertain, but usually shorter, than $t_E$ (see
Lipunov \& Popov, 1995). These authors argue, that
for non-decaying magnetic 
field $t_E$ is always longer than $t_P$. For decaying
magnetic field the situation can be different, and 
$t_P$ can be comparable to $t_E$. So we obtain here ``upper
limits'', i.e. we calculate very strong limits
on parameters $t_d$ and $\mu_b$,
which can become only wider if one takes into account the propeller stage.

 We assume exponential field decay:

\begin{equation}
\mu=\mu_0\cdot e^{-t/t_d}, \, \mu > \mu_b
\end{equation} 
where $\mu_0$ is the initial magnetic momentum 
($\mu=\frac12 B_p R_{NS}^3$, here $B_p$ -- polar magnetic field,
and $R_{NS}$-- NS radius), $t_d$ -- characteristic time
scale of the decay, and $\mu_b$ -- bottom magnetic momentum, which is 
reached in: 

\begin{equation}
t_{cr}=t_d\cdot \ln\left( \frac{\mu_0}{\mu_b} \right).
\end{equation}
After that moment magnetic field is assumed to be constant. 

In Fig. 1 we show, as an illustration, evolutionary tracks of NSs on
$P-B$-diagram for $v=10 \, {\rm km} \, {\rm s}^{-1}$ and
$n=1\, {\rm cm}^{-3}$. 
Tracks start at $t=0$, when $p=p_0=0.020 {\rm s}$ and $\mu=mu_0=10^{30} {\rm
G} \, {\rm cm}^3$, and end at 
$t=t_H=10^{10} $
yrs (for $t_d=10^{10} {\rm yrs}$, $t_d=10^9$ yrs and for constant magnetic 
field) 
or at the moment, when $p=p_E$ (for $t_d=10^8$ yrs and $t_d=10^7$ yrs). 
The line with diamonds shows $p=p_E (B)$.

As far as the accretion rate from the ISM is small (even for our parameters),
less than $\sim 10^{12} \, {\rm g} \, {\rm s}^{-1}$, 
no influence of accretion onto decay
was taken into account (see Urpin et al., 1996). 

The ejector stage lasts until the critical ejector 
period, $p_E$, is reached:

\begin{equation}
p_E=11.5 \, \mu_{30}^{1/2} n^{-1/4} v_{10}^{1/2} \, {\rm s},
\end{equation}
where  $v_{10}=\sqrt{v_p^2+v_s^2}/10 \, {\rm km} \, {\rm s}^{-1}$.
$v_p$-- spatial velocity of a NS. 
Here the sound velocity, $v_s$,
was taken into account, but as far as normally NSs spatial velocities
are higher than $10 \, {\rm km}\, {\rm s}^{-1}$ and the sound velocity
outside hot low density ISM regions is
lower than $10 \, {\rm km}\, {\rm s}^{-1}$, we have 
$\sqrt{v_p^2+v_s^2}\approx v_p$. 
$n$ is a concentration of the ISM.

Initial period should be taken to be much smaller than $p_E$. We used
$p_0=0$ s. Variations of $p_0$, if it stays much less than $p_E$, have little
influence on our results, i.e. in that case
$t_E$ is determined only by $p_E$ and history of the decay. We calculated
spin-down according to magneto--dipole formula (but other regimes are
possible, see Beskin et al. 1993 for a review):

\begin{equation}
\frac{dp}{dt}=\frac{2}{3}\frac{4\pi ^2 \mu^2}{pIc^3},
\end{equation}
where $\mu$ can be a function of time.

For our estimates we assumed constant velocity of NSs, $v$,
equal to $10 \, {\rm km} \, {\rm s}^{-1}$ and
constant ISM concentration, $n$, equal to $1 \, {\rm cm}^{-3}$. 
These conditions give us a lower limit
on $t_E$, because normally velocity is significantly higher
(fraction of slow velocity NSs is less than few percents, see Popov et al.,
1999), and ISM density is smaller than the specified values (partly because
high velocity NSs spend most of their lives in low density regions
far from the Galactic plane).

After simple algebra one can obtain a formula for $t_E$, depending upon
$t_d$, $\mu_0$, $v$, $n$ and $\mu_b$:

\begin{equation}
t_E=\left\{
 \begin{array}{cl}
 \ & \displaystyle -t_d\cdot \ln \left[ \displaystyle\frac{A}{t_d}\left(\sqrt{
\displaystyle 1+\frac{t_d^2}{A^2}}-1
\right) \right], \, t_E< t_{cr}\\
 \ & \displaystyle t_{cr}+\displaystyle A\frac{\mu_0}{\mu_b}-
\displaystyle t_d\frac{1}{2}\left(\displaystyle \frac{\mu_0}{\mu_b}\right)^2
\left(\displaystyle 1-e^{-2t_{cr}/t_d}\right), \, t_E> t_{cr}\\
 \end{array}
 \right.
\end{equation}
where coefficient $A$ is determined by the formula:

\begin{equation}
A=\frac{3Ic}{2\mu_0\sqrt{2v\dot M}}\simeq 10^{17} I_{45} \mu_{0_{30}}^{-1}
v_{10}^{-1/2}\dot M_{11} ^{-1/2} \, {\rm s},
\end{equation}
where $\dot M$ can be formally determined as a combination of intrinsic
NS's parameters, its velocity and ISM concentration using the Bondi formula
even if a NS is not on the accretor stage:

\begin{equation}
\dot M\simeq 10^{11}\,  n v_{10}^{-3} \, {\rm g}\, {\rm s}^{-1}.
\end{equation}

Results of $t_E$ calculations
for several values of $\mu_0$ and $t_d$ are shown in Fig. 2. 
Horizontal regions in the left parts of all curves appear because there
a NS reaches the critical period
$p=p_E$ before it reaches bottom magnetic momentum at 
$t=t_{cr}$, and $t_E$ is not depended upon
the rest field decay history. On the right side from the maximum on the
curves a NS reaches $\mu=\mu_b$ 
with period significantly less than $p_E (\mu_b)$,
and it takes a long time after $t=t_{cr}$
to reach $p=p_E$ (this time is longer for lower
$\mu_b$). On the left side from the maximum,
but before the horizontal region, a NS reaches $\mu=\mu_b$ with a period
close to $p_E$, and it is closer if $\mu_b$ is lower, so very quickly
after $t=t_{cr}$ a NS can spin down to $p_E$. The first point at the right
corresponds to the bottom momentum equal to initial momentum, i.e. to the
case without decay.

\begin{figure} 
\vbox{\psfig{figure=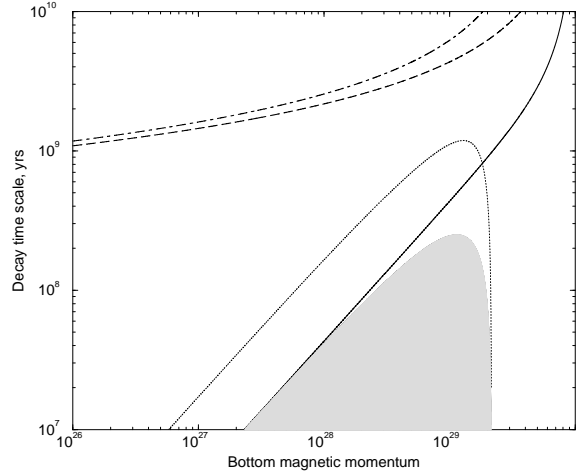,width=9.0cm,angle=-90}}
\caption{ Characteristic time scale of the magnetic field decay, $t_d$, vs.
bottom magnetic momentum, $\mu_b$.
In the filled region $t_E$ is greater than $10^{10} {\rm yrs}$.
Dashed line corresponds to $t_H=t_d\cdot \ln \left( \mu_0/\mu_b
\right)$, where $t_H=10^{10}$ years. Solid line corresponds to
$p_E(\mu_b)=p(t=t_{cr})$, where $t_{cr}=t_d\cdot \ln \left(
\mu_0/\mu_b \right)$. Both lines and filled region
are plotted for $\mu_0=10^{30} {\rm G} \, {\rm cm}^{-3}$. 
Dot-dashed line is the same as the dashed one, but for $\mu_0=5\cdot 10^{29} 
\, {\rm G} \, {\rm cm}^3$.
Dotted line is a border of the `forbidden'' region for $\mu_0=5\cdot
10^{29}  \, {\rm G} \, {\rm cm}^3$.}
\end{figure} 

\begin{figure} 
\vbox{\psfig{figure=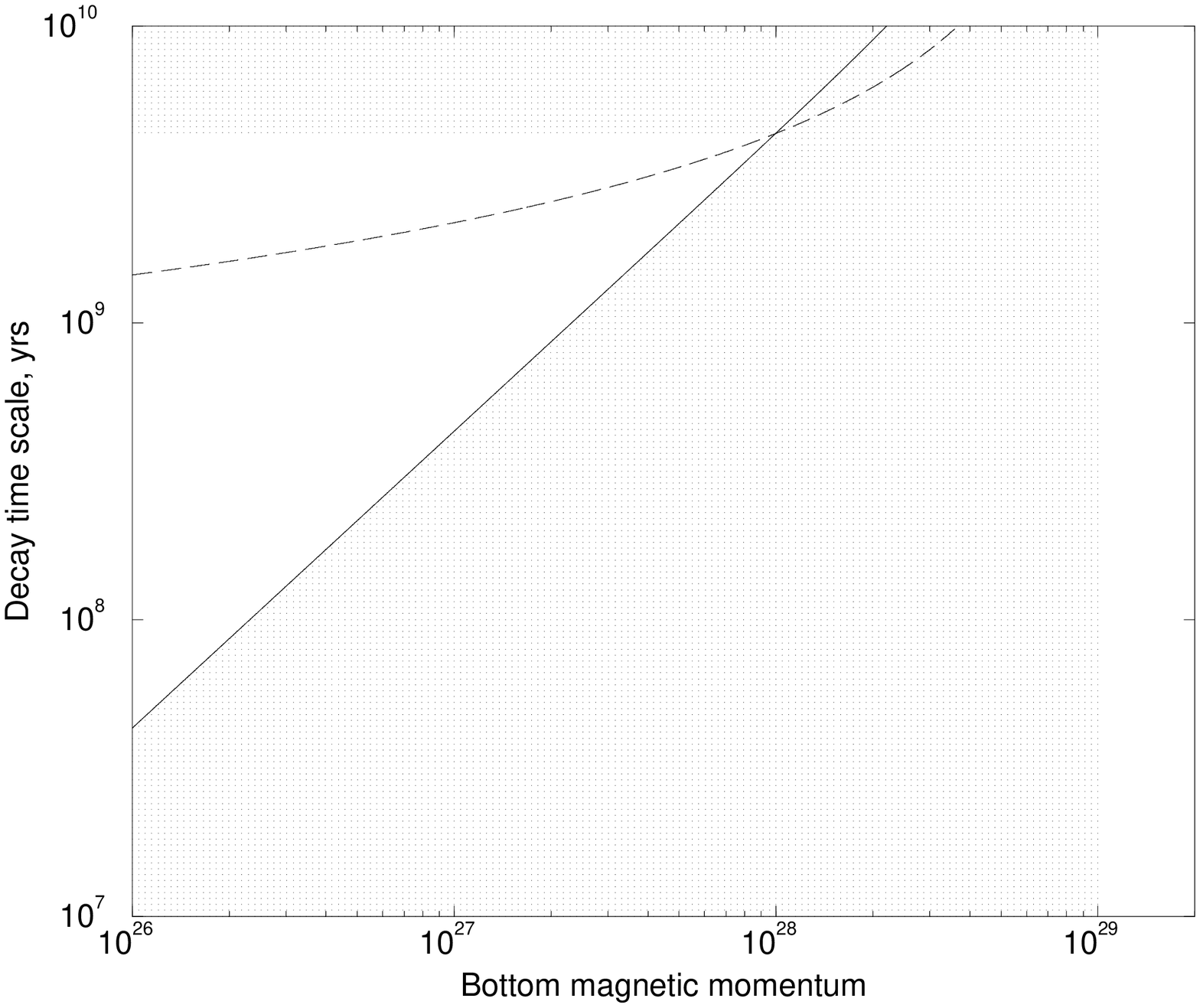,width=9.0cm,angle=-90}}
\caption{Characteristic time scale of the magnetic field decay, $t_d$, vs. bottom
magnetic momentum, $\mu_b$.
In the filled region $t_E$ is greater than $10^{10}\,  {\rm yrs}$.  
Dashed line corresponds to $t_H=t_d\cdot \ln \left( \mu_0/\mu_b   
\right)$, where $t_H=10^{10}$ yrs. Solid line corresponds to  
$p_E(\mu_b)=p(t=t_{cr})$, where $t_{cr}=t_d\cdot \ln \left(    
\mu_0/\mu_b \right)$. Both lines and region are plotted for 
$\mu_0=10^{29}\, {\rm G} \, {\rm cm}^{-3}$.}
\end{figure} 

One can see from that figure, that for some combination of
parameters $t_E$ is longer than the Hubble time. It means, that NSs never
evolve further than the ejector stage.

We argue, that as far as accreting isolated NSs are observed, combinations
of $t_d$ and $\mu_b$ for which no accreting isolated NS appear can be
excluded. We plotted it in Figs. 3 and 4.
 
Filled regions represent space of the
parameters where $t_E$ is longer than $10^{10}\, {\rm yrs}$, so in that
region a NS never come to the accretor stage, and doesn't appear as
accreting X-ray source. With the fact of
observations of accreting old isolated NSs by ROSAT this region can be called
``forbidden'' for selected
parameters of the exponential field decay (and for specified $\mu_0$). 

In the ``forbidden'' region in Fig. 3, which is plotted for
$\mu_0=10^{30}\, {\rm G} \, {\rm cm}^3$, all NSs reach the bottom field 
in a Hubble time or faster, and evolution on the late stages
of their lives goes on with the field equal to the bottom.
The left side is determined approximately by the condition:

\begin{equation}
 p_E(\mu_b)=p(t=t_{cr}).
\end{equation}
Small difference between the line, correspondent to the
condition above,
and the left side of the ``forbidden''
region appears because a NS can slightly change its period even with the
momentum $\mu=\mu_b$, but due to small value of the field angular momentum
losses are also small.
 
The right side of the region is roughly 
determined by the value of $\mu_b$, with which
a NS can reach the ejector stage with any $t_d$, i.e. 
this $\mu_b$ corresponds to the minimum value of $\mu_0$ with which a NS
reach the ejector stage without field decay.

NSs to the right from the ``forbidden'' region leave the ejector stage,
because their field cannot decay down to low values, and spin-down is
fast enough during all their lives as ejectors, because the bottom magnetic
momentum there is relatively high.
To the left from the 
``forbidden'' region the situation is different. Spin-down of NSs is very
small and they
leave the ejector stage not because of spin-down, but due to
decreasing of $p_E$, which depends upon the magnetic momentum. 

Dashed line in Fig. 3 shows, that for all interesting parameters a NS
with $\mu_0=10^{30}\, {\rm G}\,{\rm cm}^3$
reach $\mu_b$ in less than $10^{10}$ yrs.  Dot-dashed line shows the same
for $\mu_0=0.5\cdot 10^{30} \, {\rm G}\,{\rm cm}^3$.

On Fig. 3 we also show the ``forbidden'' region and the line of
reaching $\mu_b$ for $\mu_0=0.5\cdot 10^{30}\, {\rm G} \, {\rm cm}^3$.

Fig. 4 is plotted for $\mu_0=10^{29} \, {\rm G} \, {\rm cm}^3$.
For long $t_d$, $>4\cdot 10^9$ yrs, a NS again is not able to leave the
ejector stage. 
It happens because the magnetic momentum can't decrease down to small value
of $\mu$ (nearly $\mu_b$), and $p_E$ is not decreasing enough.

\section{Discussion and conclusions}

 We tried to evaluate the region of parameters,
forbidden for models of the exponential
magnetic field decay in NSs using the fact of
observations of old accreting isolated NSs in X-rays.  

 If the main fraction of NSs have nearly the same initial parameters and 
parameters of the decay, then the
intermediate values of $t_d$ ($\sim 10^7-10^8 \, {\rm yrs}$)
in combination with the intermediate values of
$\mu_b$ ($\sim 10^{28}-10^{29.5} \, {\rm G} \, {\rm cm}^3$) 
for $\mu_0=10^{30} \, {\rm G}\, {\rm cm}^3$
can be excluded, because for that sets of parameters NSs spend all
their lives on the ejector stage, never coming to the accretor stage.

As one can see in Fig. 2, for higher $\mu_0$ NSs should reach $t_E$ even
for $t_d < 10^8 $ yrs, for smaller -- the ``forbidden'' region should become
wider. Results are depended on the initial magnetic field, $\mu_0$, ISM
concentration,  $n$, and NS velocity, $v$, so
one can say, that the observed accreting isolated NSs come from, for example,
the objects with high initial magnetic field, and the rest are not visible
because they are in the forbidden region. To explore this idea in details
population synthesis of NSs for realistic distributions of $v$, $\mu_0$ and
$n$ is needed. But we can say immediately, that the idea of obtaining 
accreting old isolated NSs from initially high field objects is not very
promising, because the fraction of high field NSs can not be large (basing on
radio pulsars observations), and as far as the fraction of low velocity NSs
is not more than several percents (Popov et al. 1999) and the volume
fraction filled with relatively high density ISM is also small, accreting
old isolated NSs should come from ``typical'' population, i.e. from NSs
with $\mu_0 $ about $ 10^{30} \, {\rm G} \, {\rm cm}^3$ or less.

Actually, limits that we obtained are even stronger than 
they are in nature, because we didn't take into account,
that some significant time (in the case of field decay) a NS can spend on
the propeller stage (spin-down rate at this stage is very
uncertain, see the list of formulae, for example, in Lipunov \& Popov 1995
or Lipunov 1992). 
Calculations of this effect, and calculations for
different models of non-exponential field decay are subjects for future work.

 We cannot say anything about parameters of field decay in the case of
accretion in close binaries, because there situation is completely
different, and our results cannot be applied to millisecond radio pulsars
or other objects in close binary systems.

 We note, that there is another reason due to which
very fast decay down to small values of $\mu_b$ also can be
excluded, because it leads to huge amount of accreting isolated
NSs. This situation
is similar to ``turning-off'' the magnetic field of a NS
(i.e., quenching any magnetospheric effect on the accreting matter), 
and for any velocity and density distributions we should expect 
significantly more accreting isolated NS than we have from ROSAT observations
(of course, for high velocities X-ray sources will be very dim, but close
NSs can be observed even for velocities $\sim 100$ km s$^{-1}$).

  So, the existence of several old isolated accreting NSs, observed by ROSAT
(if it is the correct interpretation of observations),
can put important limits on the models of the magnetic field decay for
isolated (without influence of accretion, which can stimulate field decay) 
NSs, and models, from their side, should 
explain the fact of observations of $\sim 10$ accreting isolated NSs in the
solar vicinity. We cannot discuss numerous details of connection between
decay parameters and X-ray observations of isolated NSs without detailed
calculations, we just tried to show, that this connection should be taken
into account and made some illustrations of it, 
and future investigations in that field are wanted.

\begin{acknowledgements}
 We thank Monica Colpi, Denis Konenkov and Roberto Turolla 
for numerous comments on the text, suggestions and discussions. 
We also want to thank Vladimir Lipunov and Aldo Treves for advices and
attention to our work.

This work was supported by the RFBR (98-02-16801) and
the INTAS (96-0315) grants.
\end{acknowledgements}

\end{document}